\documentclass[aps,twocolumn,floats,pre]{revtex4-1}
\usepackage{graphics,graphicx,epsfig}
\usepackage{amssymb,color}
\usepackage{amsmath}
\usepackage{epsf,epstopdf,wrapfig}

\begin{document}
\title{An Integral Equation Approach to  the Dynamics of L2-3 Cortical Neurons}
\author{Richard Naud}
\affiliation{University of Ottawa}
\date{\today}

\begin{abstract}
How do neuronal populations encode time-dependent stimuli in their population firing rate? To address this question, I consider the quasi-renewal equation and the event-based expansion, two theoretical approximations proposed recently, and test these against peri-stimulus time histograms from L2-3 pyramidal cells \textit{in vitro}.  Parameters are optimized by gradient descent to best match the firing rate output given the current input. The fitting method can estimate single-neuron parameters that are normally obtained either with intracellular recordings or with individual spike trains. I find that quasi-renewal theory predicts the adapting firing rate with good precision but not the event-based expansion. Quasi-renewal predictions are equal in quality with state-of-the-art spike timing prediction methods, and does so without resorting to the indiviual spike times or the membrane potential responses.  \end{abstract}
\maketitle

\section{Introduction}
\label{sec-intro}
 Communicating with population firing rate allows for fast and reliable transfer of information \cite{Gerstner2000a,Gerstner2002c,Tchumatchenko2011a}. An important question is then: what is the mathematical function mapping stimulus to population firing rate?  The function should have parameters that can be related to the single-neuron dynamics.  

In a recent article Naud and Gerstner (2012)\nocite{Naud2012c} proposed two approximations to the dynamics of adapting populations.  The first approximation, the Event-based Moment Expansion (EME), was suggested for situations where adaptation is important but relative refractoriness weak.  It relates with previously studied models insofar as it is a generalization of the firing rate model by Benda and Herz \cite{Benda2003a}. The second approximation, the Quasi-Renewal (QR) equation, has fewer conditions but requires solving an implicit integral equation.  While Naud and Gerstner (2012) validated their approximations with Monte Carlo simulations of the spike response model, I here test these firing rate models on L2-3 pyramidal neurons recordings.  

First, I test the EME approximation and then the QR approximation.  Then, in Sect. \ref{sec-meth}, methods for estimating the single-neuron parameters from the experimental input current and output firing rate are shown. The results confirm that the QR approximation captures well the dynamics of the adapting firing rate of L2-3 pyramidal neurons.

\section{Results}
\label{sec-res}
Consider a population of $N$ neurons stimulated artificially with a current $I(t)$ at the cell body. The neurons respond with a population firing rate that can be computed by counting the fraction of neurons that fired within a small time bin centered around time $t$. In the present article, I consider a homogeneous and unconnected population. Therefore the population firing rate can be calculated with a single neuron only. Injecting $N$ times the same stimulus, the population firing rate is equivalent to the fraction of the total number of repetitions where the neuron was found active around time $t$. Tchumatchenko et al. (2011)\cite{Tchumatchenko2011a} used this approach to compute the population firing rate response $\nu(t)$ to a series of step currents. The details of the experiments are briefly described in Sect. \ref{sec-expmeth}. Let us now describe and test the EME and QR approximations.

\begin{figure}[t]
\centering
\includegraphics[width=.49\textwidth]{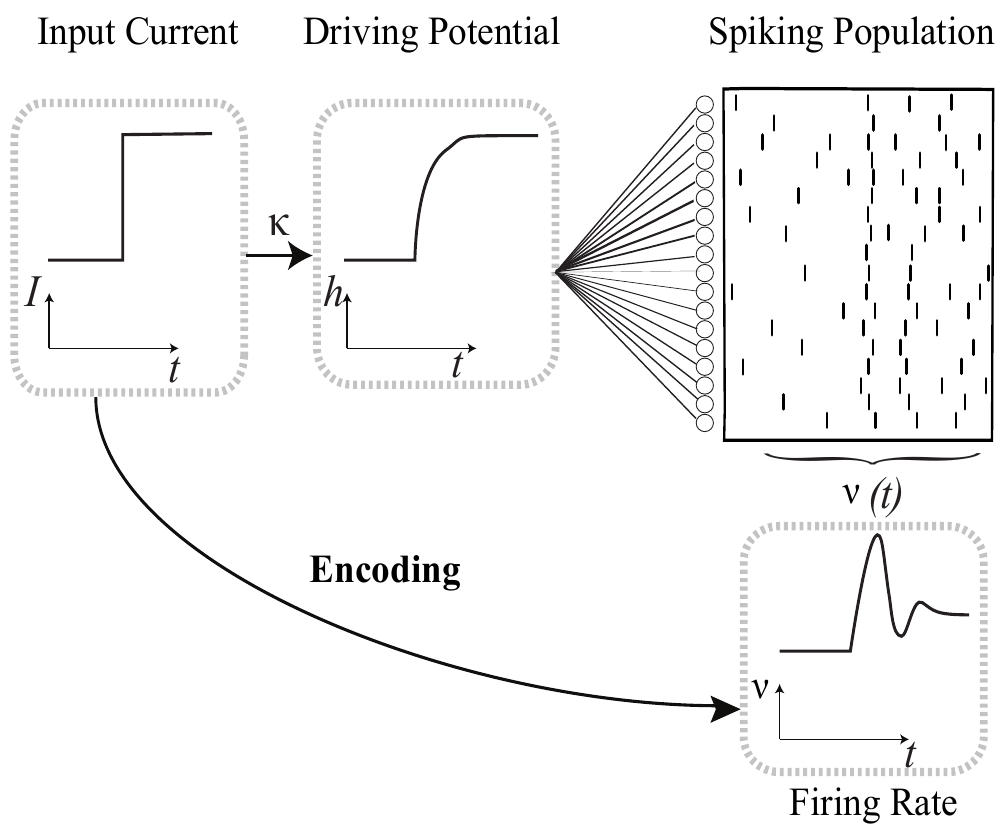}
\caption{Encoding step input changes in the firing rate of an unconnected neuronal population, a schematic representation of the neuron models. The time-dependent current, $I$, drives change in the subthreshold membrane potential, $h$, of the model the neuron. The relation between input current and driving potential is a linear convolution described by the filter $\kappa$. The same driving potential affects all neurons in the unconnected population. Each neuron fires stochastically with a probability that follows the distance to an adapting threshold. The spikes (black ticks) are then averaged across the population to give a time-dependent firing rate $\nu$.  Adaptation and refractoriness causes the time dependent firing rate to differ considerably from the time-dependent input. }
\label{fig-schem}
\end{figure}

\subsection{EME Approximation}
\label{sec-emeres}
\begin{figure*}[t]
\centering
\parbox{.03\textwidth}{\bf \large A\\ \vspace{3.5cm}  B \\ \vspace{3cm}  C \vspace{3.0cm}}
\parbox{.96\textwidth}{
\includegraphics[width=.9\textwidth]{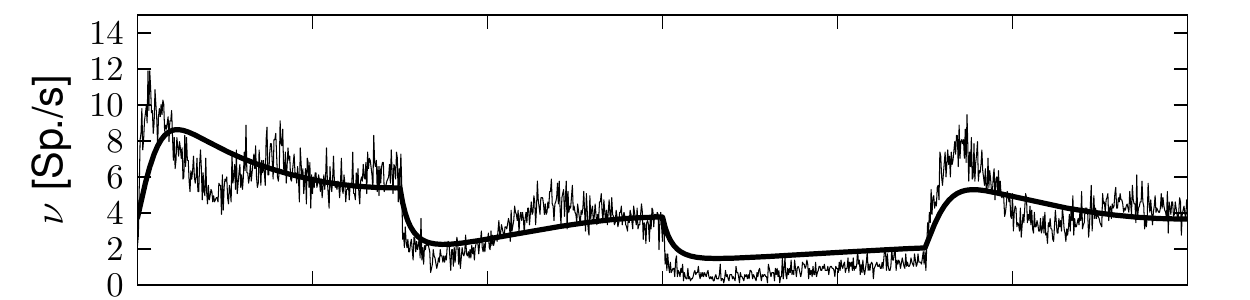}
\includegraphics[width=.9\textwidth]{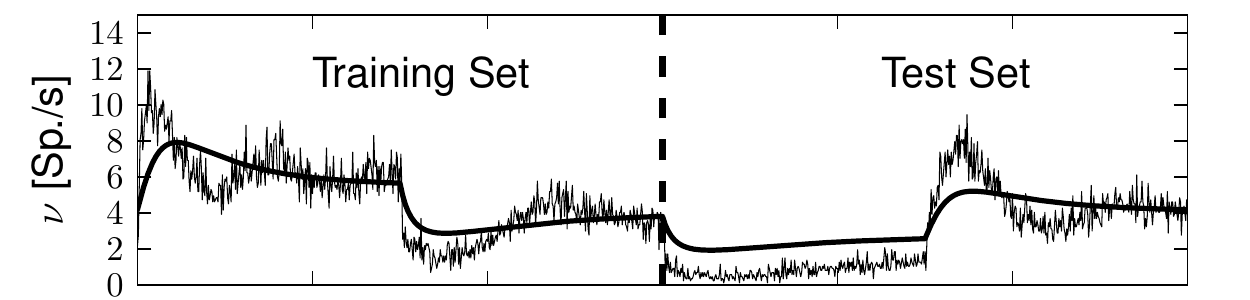}
\includegraphics[width=.9\textwidth]{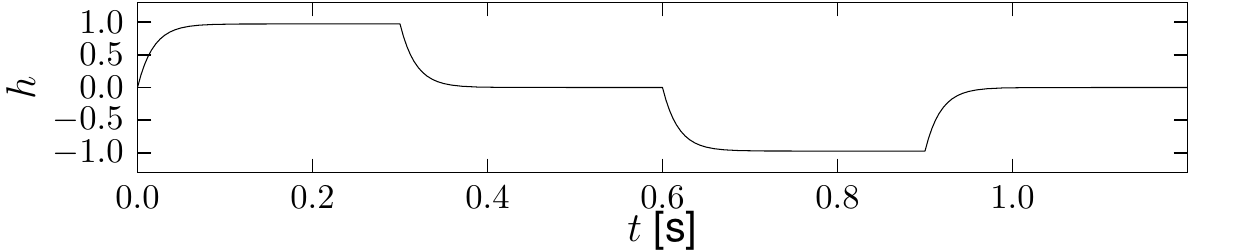}
}
\caption{Firing rate prediction using the EME approximation. {\bf A}: Observed firing rate (thin black line) and model firing rate (thick black line) fitted on the whole data set. {\bf B}: Same as A but using only the first 600~ms for parameter estimation. {\bf C}: driving potential $h(t)$.  }
\label{fig-eme}
\end{figure*}

The Event-based Moment Expansion assumes the coupling between individual spikes is sufficiently small such that the firing rate follows 
\begin{equation}
\nu(t) = \nu_0 \exp \left( {\rm h}(t) + \int_{-\infty}^t \left( e^{\eta(t-z)}-1\right) \nu(z)dz \right)
\label{eq-eme}
\end{equation}
where ${\rm h}(t)$ is the driving potential and $\eta(t)$ the spike after-potential.  The parameter $\nu_0$ is a scaling constant that can be set arbitrarily to 1~kHz.

Sect. \ref{sec-meth} describes how the parameters defining ${\rm h}(t)$ and $\eta(t)$ can be determined from the knowledge of the input current and the observed firing rate. The fitting method uses a multi-linear regression to arrive to an initial estimate of the parameters defining $\eta(t)$. In a second step, the root mean square error (RMSE) between modeled and observed firing rate is used to perform a gradient descent.

 As a first test to the theory, we use the full dataset to fit the parameters. The data set consists of 1.2 seconds of current injection and the observed firing rate (Sect. \ref{sec-expmeth}).  Using the fitted parameters,  Eq. \ref{eq-eme} is simulated to produce the firing rate shown in Fig. \ref{fig-eme}A. The model reproduces the data grossly: the root mean square error (RMSE) was 1.3 Hz and the variance explained (labeled $M_D$, see Sect. \ref{sec-ememeth})was 53\%. 

To avoid overfitting, we separated the full 1.2-second dataset in two sets. The first 0.6~s were used to extract the parameters and the remaining 0.6 seconds were used to test the model performance. In such a prediction task (Fig. \ref{fig-eme}B) the variance explained dropped from 53\% to 38\%, indicating the EME approximation cannot predict the firing rate response accurately.  The RMSE and $M_D$ for training and test sets are summarized in Table~\ref{table1}.

\subsection{QR Approximation}
\label{sec-qrres}
 
\begin{table}
\begin{tabular}{c|c|c|c|c}
Model & Test RMSE  & Train. RMSE& Test $M_D$ & Train. $M_D$ \\ \hline
QR(1 ms) & 1.07 Hz & 0.92 Hz & 70 \% & 79 \% \\ 
 QR(8 ms) & 0.96 Hz & 0.63 Hz & 75 \% & 90 \% \\ 
 EME(1 ms) & 1.37 Hz & 1.30 Hz & 38 \% & 53 \% \\ 
 EME(8 ms) & 1.29 Hz & 1.10 Hz & 43 \% & 64 \% 
\end{tabular}
\caption{Performance of EME and QR approximations in terms of the RMSE and $M_D$ evaluated with $\Delta t =1$~ms or $\Delta t=8$~ms.}\label{table1}
\end{table}

\begin{figure*}[t]
\centering
\parbox{.03\textwidth}{\bf \large A\\ \vspace{4.5cm}  B \\ \vspace{3.5cm}  C \vspace{3.0cm}}
\parbox{.96\textwidth}{
\includegraphics[width=.9\textwidth]{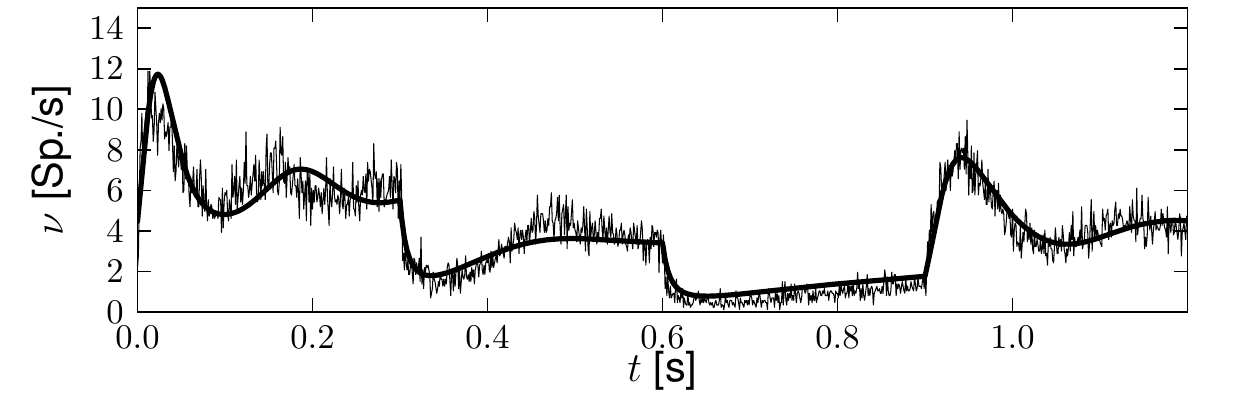}
\includegraphics[width=.9\textwidth]{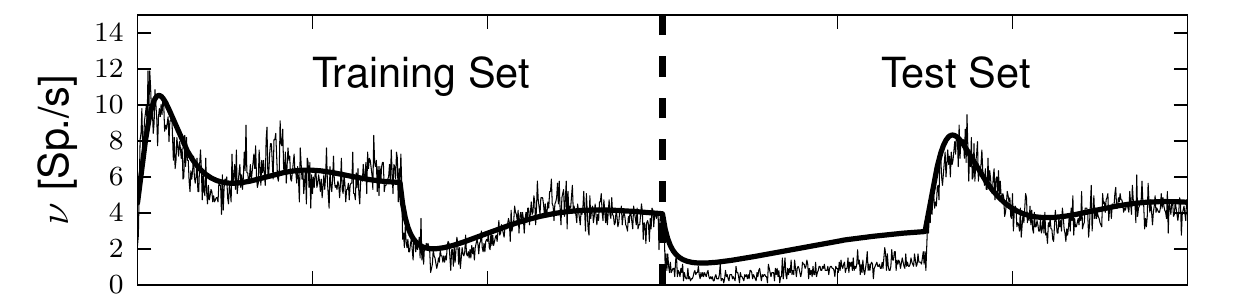}
\includegraphics[width=.9\textwidth]{h_Tchum.pdf}
}
\caption{Firing rate prediction using the QR approximation. {\bf A}: Observed firing rate (thin black line) and model firing rate (thick black line) fitted on the whole data set. {\bf B}: Same as A but using only the first 600~ms for parameter estimation. {\bf C}: driving potential $h(t)$.  }
\label{fig-qr}
\end{figure*}

Quasi-renewal theory describes the dynamics of neurons by defining a survivor function $s_A(t|\hat{t})$. The survivor function describes the probability of not firing at time $t$ given a previous spike at time $\hat{t}$.  Classical renewal theory concerns stationary input and survivor functions\cite{Cox1962a}, it cannot account for step changes in input. If one is to follow time-dependent renewal theory \cite{Gerstner1995c}, the survivor function would only depend on the input to the neurons before time $t$.  But in quasi-renewal theory, adaptation makes neurons less likely to spike given recent activity. Therefore, the survivor function also depends on the previous firing rate history. Naud and Gerstner (2012) derived the following survivor function:
\begin{equation}\label{eq-surv}
s_{\nu,I}(t|\hat{t}) = \exp\left( - \int_{-\infty}^t \lambda(t|\hat{t}) dx \right)
\end{equation}
where $\lambda(t|\hat{t})$ is the conditional probability intensity of emitting a spike at time $t$ given a previous spike at time $\hat{t}$. The instantaneous firing intensity becomes:
\begin{eqnarray}\label{eq-lam}
\lambda(t|\hat{t}) =&\\
\lambda_0 \exp& \left({\rm h}(t) -\eta(t-\hat{t}) - \int_{-\infty}^{\hat{t}} (e^{\eta(t-z)}-1)\nu(z) dz \right), \nonumber 
\end{eqnarray}
where $\lambda_0$ is a scaling factor that can arbitrarily be set to 1 ms$^{-1}$.  

Using the survivor function, the conservation equation 
\begin{equation}\label{eq-qr}
1 = \int_{-\infty}^t s_{\nu,I}(t|\hat{t}) \nu(\hat{t}) d\hat{t}
\end{equation}
ensures that all neurons have emitted their last spike at some time in the past.  Eq. \ref{eq-qr} can be used with the definitions in Eqs. \ref{eq-surv} and \ref{eq-lam} to determine $\nu(t)$ numerically. 

The single-neuron parameters implicitly defined in $\eta(t)$ and $h(t)$ appear recursively in the nested integrals of Eqs. \ref{eq-surv} - \ref{eq-qr}.  I did not succeed in finding a convex optimization method. Instead I used the standard non-linear fitting procedure of Levenberg-Marquardt \cite{Marquardt1963a} with repeated random initializations.  Fitting on the whole data set (Fig. \ref{fig-qr}A) nevertheless yielded a good match (RMSE = 0.82~Hz, $M_D$ = 87\%).  When training on the first 600~ms of recordings and testing on the remaining 600~ms (Fig. \ref{fig-qr}B), the method showed little overfitting as can be seen from the small gap in performance between training and test sets in Table \ref{table1}.

\begin{figure}
\centering
\includegraphics[width=.49\textwidth]{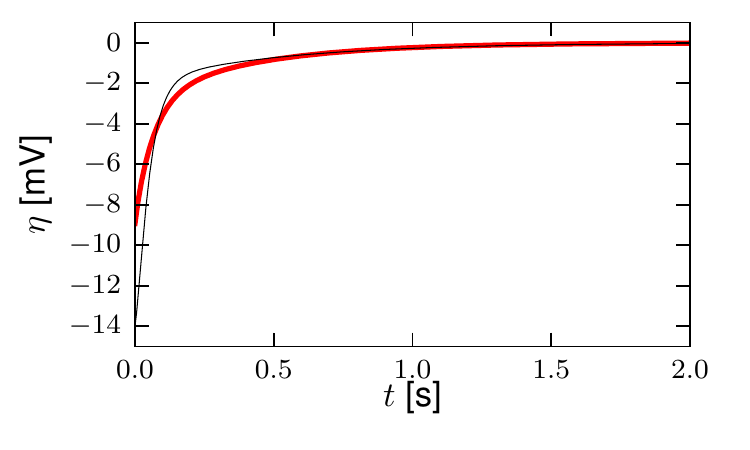}
\caption{Fitted single neuron parameters match intra-cellular measurements. The kernel $\eta(t)$ fitted on the firing rate data with the QR approximation (red line) compares well with the equivalent kernel obtained from intracellular recordings (black line, obtained from Mensi {\it et al.} \cite{Mensi2012a}).}
\label{QRfit}
\end{figure}

Finally, I compared the single neuron parameters obtained from the fit on the firing rate with those measured in intracellular recordings. The effective spike after potential was obtained from intracellular recordings in Mensi {\it et al.}\cite{Mensi2012a} by combining the moving threshold with the spike after-currents. This effective spike after-potential does not vary considerably from cell to cell\cite{Mensi2012a}. I used the spike-after potential averaged across all cells to compare with the spike after potential fitted on a single neuron. Figure \ref{QRfit} compares the spike after potentials $\eta(t)$. The present results matched the measured $\eta(t)$ at times since the last spike greater than 0.5 seconds. There are discrepancies in the early refractory period, in particular there is a underestimation of the early ($t<50$~ms) spike after potential. I conclude that the single neuron parameters estimated from the firing rate are consistent with the real single neuron parameters. Further work will be required to perform a quantitative assessment.

\section{Methods}
\label{sec-meth}
First, the experimental methods are described then the fitting methods and finally the analysis methods. The fitting methods are in three parts. First, I consider the assumptions required to estimate the driving potential from the injected current and other typical L2-3 pyramidal cell properties (Sect. \ref{sec-hmeth}). Then I describe how parameter estimates were initialized and then how the best set of parameters is determined.  The same method was used for the EME (Sect. \ref{sec-ememeth}) and QR (Sect. \ref{sec-qrmeth}) approximations.  The methods for evaluating the model performance are described in Sect. \ref{sec-analyse}.

For numerical methods on evaluating the self-consistent equation (Eq. \ref{eq-qr}) see Naud and Gerstner (2012)\nocite{Naud2012a}. Code is available on the author's website.

\subsection{Electrophysiological Data}
\label{sec-expmeth}
In vitro recordings from L2-3 pyramidal neurons were graciously shared by T. Tchumatchenko. The methods were described in details in the original work \cite{Tchumatchenko2011a} and in earlier work in this direction \cite{Volgushev2000a}. Briefly, Wistar rats (P21-P28; Harlan) were anesthetized and then decapitated for their brains to be removed rapidly. A single hemisphere was laid on an agar block and then sliced in the sagittal axis with a vibratome.  The slices containing the visual cortex were placed in an incubator for an hour of recovery. Then, in a recording chamber, slices were perfused with a solution containing (in mM) 125 NaCl, 2.5 KCl, 2 CaCl$_2$, 1 MgCl$_2$, 1.25 NaH$_2$P)$_4$, 25 NaHC)$_3$, and 25 D-glucose, bubbled with 95\% O$_2$ and 5\% CO$_2$. Temperature was kept between 28 and 32 degrees Celsius.  Whole-cell recordings using patch electrodes were made from layer 2/3 pyramidal neurons.  Current injections were made in batches of 46 s and were interleaved with 60-100 s recovery periods. The membrane potential was recorded with a sampling frequency of 10~kHz such that the bin size was 0.1~ms.

The current injection consisted of 4 segments of 300-ms duration. The current template $I(t)$ was made of a series of steps:
\begin{equation}
I_1(t) = \left\{ \begin{split}
1 &\, &{\rm for }\;0&< t<300  \; {\rm ms}\\
0&\, &{\rm for }\;300&< t<600  \; {\rm ms}\\
-1 &\, &{\rm for }\;600&< t<900  \; {\rm ms}\\
0 &\, &{\rm for }\;900&< t<1200  \; {\rm ms}\\
\end{split}
\right. 
\end{equation}
A shifted, and noisy version of $I(t)$ was repeatedly injected through the patch electrode in the L2-3 pyramidal neurons:
\begin{equation}
I(t) = A_1 I_1({\rm mod}(t,T\Delta t)) +I_0 + \sigma \epsilon(t)
\end{equation}
where $A_1$ and $\sigma$ are scaling factors, $I_0$ determines the baseline current, and $\epsilon(t)$ is an Ornstein-Uhlenbeck process with zero mean, unit variance and correlation time of 5 ms. The noise term models the input fluctuations to be expected for L2-3 pyramidal neurons in a balanced excitation-inhibition regime.  Repeated injection allowed an estimation of the the instantaneous firing rate $\nu^{({\rm obs})}(t)$.  A total of $N=$~8664 repetitions were used to compute the peri-stimulus time histogram (PSTH). Averaging over all recorded spike time $\hat{t}_i$ in terms of their phase with respect to the step stimulus $I_1(t)$ gives the firing rate:
\begin{equation}
\nu^{({\rm obs})}(t) = \frac{1}{N T\Delta t}\sum_i{\rm mod}\left(\hat{t}_i,T\Delta t \right)
\end{equation} 
where $T$ is the total number of time steps and $\Delta t$ is the binsize such that $T \Delta t=$~1.2 s.  The time step was chosen to 1~ms for parameter estimation and evaluating the goodness-of-fit. We also evaluated the goodness-of-fit with $\Delta t=8$~ms to conform with the typical precision in spike time prediction \cite{Gerstner2009a,Mensi2012a}.

\subsection{Estimate of the Driving Potential}
\label{sec-hmeth}
The driving current $I_1(t)$ causes changes in the membrane potential. Assuming that the membrane time constant is $\tau_m=18$~ms as in previous measurements in L2-3 pyramidal neurons \cite{Mensi2012a}, we can obtain an approximation of the driving potential:
\begin{equation}
h(t) = \frac{1}{ \tau_m} \int e^{-(t-z)/\tau_m} I_1(z)dz. \label{eq-hqr}
\end{equation}
The driving potential is formed by an exponential filter of the input current.
Eq. \ref{eq-hqr} remains an approximation since it assumes that a single exponential is sufficient to account for the subthreshold dynamics and that this single exponential has its time constant fixed to 18~ms.  The driving potential $h(t)$ remains to be scaled and offseted to form the driving potential ${\rm h}(t)$ in Eq. \ref{eq-eme} and Eq. \ref{eq-lam}: ${\rm h}(t) = c h(t) + h_0$. The scaling of the driving potential $c$ relates to the capacitance of the cell body and is a parameter to be fitted.

\subsection{Initialization Procedure}
\label{sec-ememeth}
To initialize the parameters, I use the convex, linear regression problem of estimating the parameters that best describe the logarithm of the firing rate in the EME approximation.
Consider the observable $y_j$ made of the logarithm of the firing rate at time $j\Delta t$: $y_j = \log \nu_j^{\rm (obs)}$. The EME approximation (Eq. \ref{eq-eme}) in discrete time becomes:
\begin{equation}
y_j  = h_0 + c h_j + \sum_{k=0}^K \xi_k \nu_{j-k}. \label{eq-emereg}
\end{equation}
Where $\xi_j = (e^{\eta(j\Delta t)}-1)\Delta t$ weighs the adapting effects of past activity and $h_j$ represents the driving potential discretized on the same grid as $\nu_j$.  To formulate Eq. \ref{eq-emereg} in a linear regression problem,a parameterization of $\xi_j$ is introduced. Here I used $p=6$ exponential bases with log-spaced time constants $\tau_i$.  Thus using $\xi_j = \sum_{i=1}^{p-2}a_i e^{j\Delta t /\tau_i}$ and $h_j = h(j\Delta t)$  casts Eq. \ref{eq-emereg} in matrix form:
\begin{equation}
\mathbf{y} = X\mathbf{\theta}
\end{equation}
where $\mathbf{y}$ is a  column vector of length $K$, $X$ is a $K \times p$ matrix and $\theta$ is a column vector of length $p$ containing the parameters to be determined.  I constructed $X$ such that the first column was uniformly filled with ones, the second column contained the raw input estimate $h(t)$ and the remaining $p$=6 columns in $X$ contained the observed activity filtered with an exponential filter having $p$ log-spaced time-constants $\tau_i$. Constructed this way, the vector of parameters is $\mathbf{\theta}=\{ h_0, \,c, \, a_1, \, a_2, \, ... ,\, a_{p} \}^T$. 

The set of parameters that minimizes the mean-square error in $y_j$ is then \cite{Weisberg2005a}:
\begin{equation}
\hat{\theta} = (X^TX)^{-1}X^T \mathbf{y}.
\end{equation}
Typically, parameters obtained this way yield unrealistic $\xi(t)$ kernels with segments greater than zero. Such kernels give runaway numerical solutions to either Eq. \ref{eq-eme} or Eq. \ref{eq-qr}.  It is, however, an efficient method to obtain an initial guess of the parameters. The initial guess is formed by replacing all positive parameters $a_1$, ... $a_p$ by zero.  Note that for this initialization procedure to work, the number of parameters $p$ must be sufficiently small to prevent overfitting from creating matching pairs of exponentials with opposite polarity.

\subsection{Parameter Estimation}
\label{sec-qrmeth}
Parameter estimation was performed using a gradient descent of the root-mean squared error (RMSE, see Sect. \ref{sec-analyse}) between model and observed firing rate.  Using the initial guess for the set of parameters $\mathbf{\hat{\theta}}_1$ the model firing rate is calculated using either Eq. \ref{eq-eme} for the EME approximation or \ref{eq-qr} for the QR approximation.  This firing rate was used to calculate the RMSE. Then the estimate of $\theta$ is modified following standard Levenberg-Marquardt least-squares algorithm. The best estimate of $\theta$ is then recorded before reinstating the gradient descent with initial guess $\mathbf{\hat{\theta}}_2 = \mathbf{\hat{\theta}}_1 + \delta \theta$ where $\delta \theta$ is a vector of random numbers drawn from a Gaussian distribution with standard deviation of 0.5. The initialization and optimization steps are repeated $n$ times. All the $n=20$ initializations yielded similar parameters but one, which had a marginally large RMSE of 4~Hz.  Therefore this method, although not convex, yields a robust and accurate estimate of the parameters.

\subsection{Analysis Methods}
\label{sec-analyse}
The root mean square error between the model firing rate $\nu^{\rm (mod)}(t)$ and the observed firing rate $\nu^{\rm (obs)}(t)$ is :
\begin{equation}
{\rm RMSE} = \sqrt{\frac{1}{T_\psi}\int_\psi \left( \nu^{\rm (obs)}(t) - \nu^{\rm (mod)}(t) \right)^2 dt } 
\end{equation}
where $\psi$ denotes the ensembles of times on which the RMSE is evaluated and $T_\psi = \int_\psi dt$ the total amount of time it spans. We mainly considered two subset of the entire experiment which we refer to training and test sets, defined in Sect. \ref{sec-res}.

In order to compare with other published work on predicting spike times, we also computed the variance explained\cite{Naud2011a}
\begin{equation}
M_D = 1- \frac{2\;{\rm RMSE} ^2}{{\rm Var}[\nu^{\rm (mod)}] + {\rm Var}[\nu^{\rm (obs)}]} \label{eq-md}.
\end{equation}
Where, implicitly, the mean squared error and the variances in Eq. \ref{eq-md} are evaluated on the same subset of time, $\psi$. Such a measure of explained variance was used to evaluate model performance in the international spike time prediction competition \cite{Gerstner2009a,Naud2011a} as well as in other spike-time prediction scenarios \cite{Pillow2008a}.

\section{Discussion}
Most spike time metrics can be cast in a comparison of instantaneous firing rates  such as $M_D$ \cite{Naud2011a}.  
This measure was used in previous studies to determine how various models predicted the spike times.  Mensi \textit{et al.} (2012)\nocite{Mensi2012a} used a spike response model to predict spike times of L2/3 pyramidal neurons with $M_D=0.81\pm$0.04. In L5 pyramidal neurons the international spike timing prediction competition\cite{Gerstner2009a,Naud2011a} concluded that the state-of-the-art single-neuron model achieved $M_D=0.74$ on average.   These studies typically use a smoothing parameter that is equivalent to binning the firing rate with bins of $\Delta t=8$~ms. At this level of precision, the QR approximation could predict $M_D=0.75$. Therefore, the firing rate prediction of the QR approximation is comparable to the state-of-the-art spike time prediction of model fitted on intracellular recordings. 

Limitiations of the methods and results present here call for further work in order to assess the validity of QR theory.  Parameter estimation was performed here with a very small training set.  Only 600~ms were used, the typical training set consists of at least 10~seconds\cite{Mensi2012a}.  The restricted size of the data set also prevents further analysis of the fitting methods.

 Another important assumption to verify in additional work is the assumption of homogeneity. The validity of the QR approximation for an heterogenous population of neurons remains to be determined.

  Finally, the predictive potential of other firing models should be assessed. For instance the moving threshold models\cite{Tchumatchenko2011b} or those based on the Fokker-Planck equation\cite{Muller2007a,Toyoizumi2009a}. Biophysical processes not taken into account by the QR approach could also play a role. Indeed, the long effect of spike after potential can modify the firing rate response to periodic input\cite{Pozzorini2013a}. Another example is the coupling between the moving threshold and subthreshold membrane potential \cite{Azouz2000a}.

\acknowledgements
Thanks to T. Tchumatchenko for sharing the data and to W. Gerstner for helpful suggestions.

\bibliographystyle{h-physrev}

\end{document}